# Triple Vector Boson Production at the LHC

**Dan Green**
**Fermilab - CMS Dept.**
(dgreen@fnal.gov)

**Introduction**

The measurement of the interaction among vector bosons is crucial to understanding electroweak symmetry breaking. Indeed, data from LEP-II have already determined triple vector boson couplings and a first measurement of quartic couplings has been made [1]. It is important to explore what additional information can be supplied by future LHC experiments. One technique is to use "tag" jets to identify vector boson fusion processes leading to two vector bosons and two tag jets in the final state. [2]

However, the necessity of two quarks simultaneously radiating a W boson means that the cross section is rather small. Another approach is to study the production of three vector bosons in the final state which are produced by the Drell- Yan mechanism. There is a clear tradeoff, in that valence-valence processes are not allowed, which limits the mass range which can be explored using this mechanism, although the cross sections are substantially larger than those for vector boson fusion.

**Triple Vector Boson Production**

A Z boson decaying into lepton pairs is required in the final state in order to allow for easy triggering at the LHC and to insure a clean sample of Z plus four jets. Possible final states are ZWW, ZWZ and ZZZ. Assuming that the second and third bosons are reconstructed from their decay into quark pairs, the W and Z will not be easily resolved given the achievable dijet mass resolution. Therefore, the final states of interest contain a Z which decays into leptons and four jets arising for the quark decays of the other two vector bosons.

Each of these final states has a contribution from quartic coupling of the vector bosons. In order to assess the relative production cross sections the COMPHEP program [3] was first used. The process with the largest cross section is $W^+ZW^-$ production. Guided by that initial exploration, this specific reaction can be used in a more detailed Monte Carlo study at a later stage

**COMPHEP Results**

The Feynman diagrams for ZWW production by way of u quark-antiquark annihilation are shown in Fig.1. There are sixteen distinct diagrams, two with quartic ($WWZ\gamma$ or $WWZZ$) vertices.

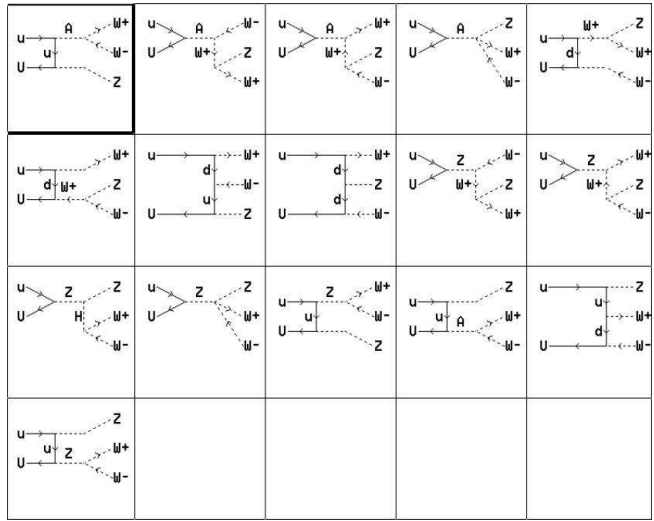

Figure 1: Feynman diagrams for Drell-Yan production of the ZWW final state. Note the contributions from γWWZ and ZWWZ quartic couplings.

The cross section for the specific Drell-Yan process as a function of C.M. energy is shown in Fig.2.

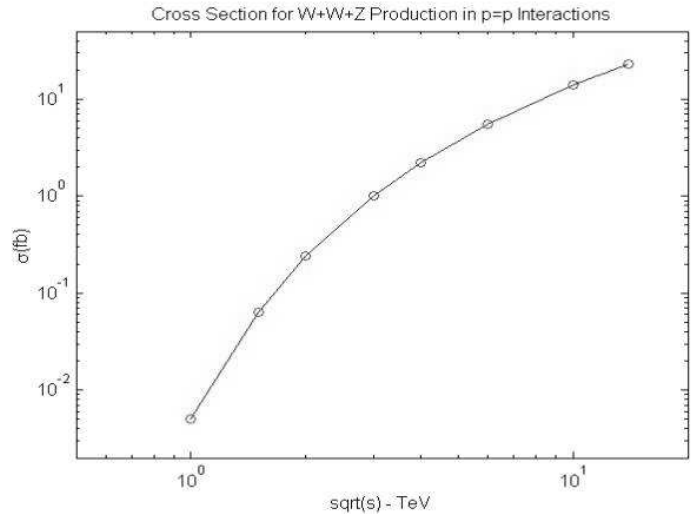

Figure 2: Cross section for u quark-antiquark annihilation and production of the WWZ final state as a function of C.M. energy.

There is evidently an increase of a factor of about one hundred as the C.M. energy rises from the Tevatron at 2 TeV to the LHC at 14 TeV. Given that large factor, and the rather small cross section, no further study is made except at LHC energies.

The cross section for the process $u + \bar{u} \rightarrow W^+ + W^- + Z$ at the LHC in p-p collisions is about 27 fb. The process initiated by d quark-antiquark annihilation has a cross section of about 22 fb. Summing $u + \bar{u}, d + \bar{d}, \bar{u} + u, \bar{d} + d$, a cross section of ~ 100 fb is expected at the LHC for WWZ production.

The ZZZ final state has a cross section about one order of magnitude smaller, and will be ignored. The ZZW final state arises from the $u + \bar{d}$ initial state and has a cross section of about 11 fb at the LHC. Essentially the process is off shell Drell-Yan production of a W which then radiates two Z bosons. Therefore, the complete cross section at the LHC for ZVV production is $(u+\bar{d}, d+\bar{u}, \bar{u}+d, \bar{d}+u)$ about 150 fb, where V = W or Z which is an order of magnitude larger than the cross section for W-W fusion production of Z-Z pairs [2].

Some of the kinematic variables for the vector bosons in the final state are shown in Fig.3 and Fig. 4. The mass of the WW pair is quite low while the mass of the WWZ system peaks at about 400 GeV and extends up to a scale of ~ 1 TeV although the mean value of the u valence quark is limited to rather low values, $<x_u>$ = 0.18. The mean WW mass is ~ 380 GeV while the mean WWZ mass is ~ 700 GeV, due to the long high mass tails of the distributions.

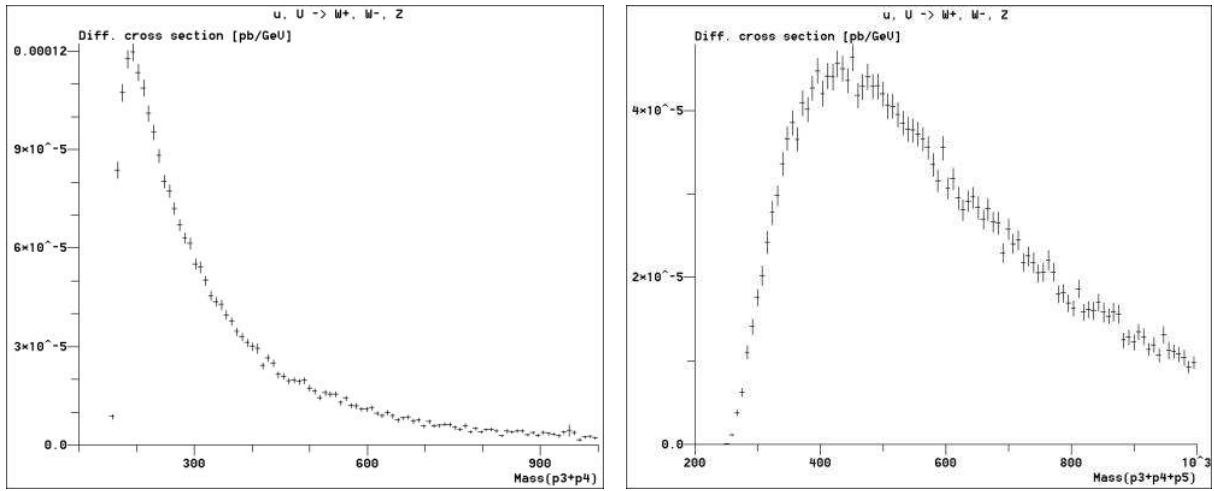

Figure 3: Mass distribution of the WW and WWZ systems in Drell-Yan production at the LHC.

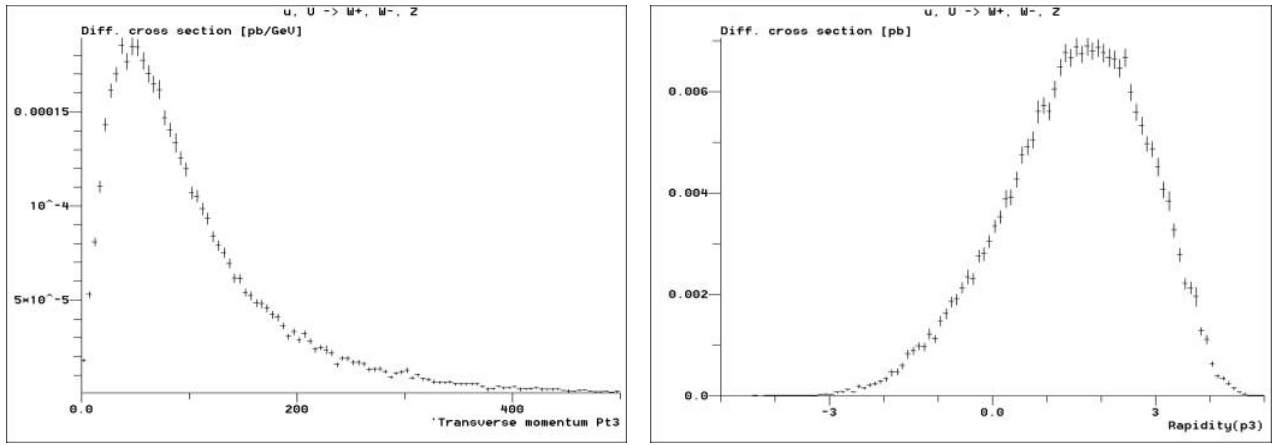

Figure 4: Transverse momentum and rapidity distribution of the W boson in the WWZ final state from Drell-Yan production at the LHC

The fairly low values of transverse momentum and the rather large values of rapidity mean that the efficiency for geometric detection and triggering are not expected to be extremely large. The mean transverse momentum of the W and the Z is ~ 125 GeV, while the mean rapidity of the W is ~ 1.5, while the Z mean rapidity is ~ 0.5. Therefore, we also expect that the geometric efficiency of the Z will be higher than that of the W. The correlation between the rapidities of each pair of vector bosons is observed to be small.

**PYTHIA Results**

The WWZ events which were generated by COMPHEP were used as input to Pythia [4], where the Z was required to decay leptonically while both W were required to decay into quark pairs. Assuming a cross section of 150 fb, then in one LHC year ( 100 $fb^{-1}$), there are ~ 450 events where the Z decays into electrons or muons and the W decays into quark pairs. Typically these events would appear in the dilepton trigger stream of a generic LHC experiment.

The quark jets are hadronized in the Monte Carlo program. The average number of hadrons plus photons in the final state is about 250. The leptons are identified and the invariant mass checked to find the pair which are the decay products of the Z boson. Cuts are made on the leptons which approximate the geometric acceptance of a generic detector and the trigger thresholds for dilepton candidates. The magnitude of rapidity must be < 2.5 and the transverse momentum of both leptons in the dilepton trigger is > 5 GeV. The dilepton efficiency for geometric and trigger cuts is about 72 %, or about 85 % per lepton in the absence of correlations. These efficiencies are higher than those found below for the jets, as expected from the prior discussion.

The final state hadrons are examined in a search for jets using a simple cone algorithm. The seed particle has a transverse momentum > 3 GeV and a cone of radius 0.3 is used. This small cone size is used because the W bosons possess a substantial momentum causing the quark (jet) – quark (jet) opening angle to be fairly small and the two jets must be separated by an opening angle of at least two cone radii. The average number of found jets is 7.2. The four highest transverse momentum jets are selected for further analysis.

The dijet mass is shown in Fig.5. Note that there is no energy smearing applied here due to finite detector resolution. The W width is entirely due to out of cone jet fluctuations around the narrow cone boundary. A wide cut of +- 20 GeV around the W mass is applied. The correlation between dijet masses is shown in Fig.6. There are three entries of two mass pairs per event, which fills in the scatterplot with combinatoric background .Nevertheless, a clear enhancement of W pairs decaying into quark dijets is evident.

The overall efficiency to find the correct four jets and reconstruct the mass of the W within the cut window is about 47%, or about 68% per W assuming no correlations. Clearly, more work is needed in the future in order to optimize the reconstruction efficiency by working on the jet algorithm.

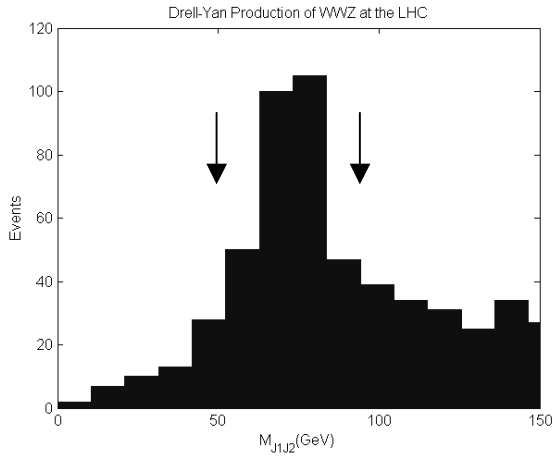

Figure 5: Dijet mass distribution for the six pairs possible using the four highest transverse momentum jets in ZZW events. The arrows indicate acceptance limits.

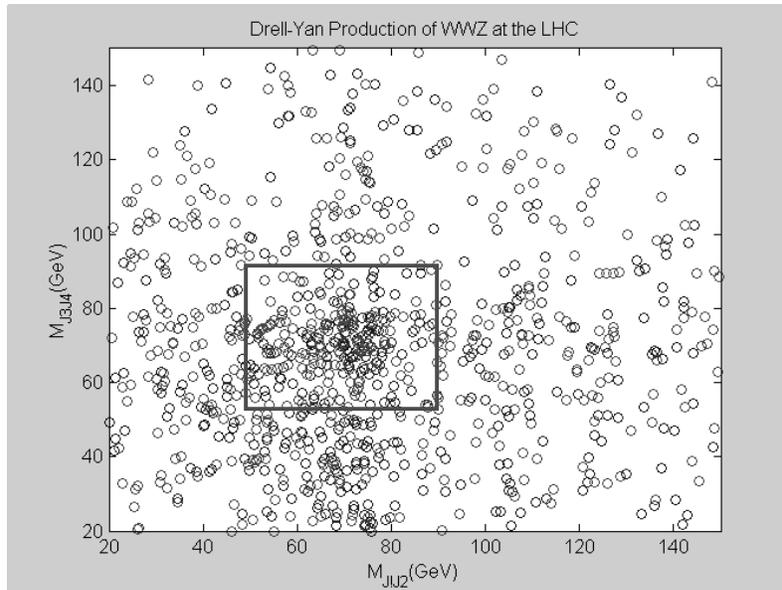

Figure 6: Scatterplot of the pair of dijets in the event, where there are three possible combinations of the four jets in the event. The box indicates the joint acceptance window for the W mass.

Cuts on transverse momentum and rapidity were imposed to simulate the geometric and trigger acceptance of a generic LHC detector for jets. The transverse momentum of each of the four jets must be > 20 GeV and the magnitude of the rapidity must be < 5 for a jet. The efficiency per dijet is about 70%, while the overall WW efficiency is 49%.

Overall the event efficiency to strike the appropriate detector with two leptons and four jets, trigger the detector, and have the mass reconstructed as a Z dilepton and two W dijets is about 16%. The number of reconstructed events is then approximately 72 for one LHC year at design luminosity or 100 fb$^{-1}$. Clearly, a large effort is needed to improve on the reconstruction efficiency.